\def\beg{\begin{equation}}
\def\eeq{\end{equation}}
\def\begarr{\begin{eqnarray}}

\documentstyle[12pt]{article}
\begin{document}
\baselineskip20pt

\begin{center}
{\bf Effect of the electric field on a superconducting powder}\\
\vskip0.5cm

 Keshav N. Shrivastava*\\
School of Physics, University of Hyderabad, Hyderabad 500 046, India.
\end{center}
\vskip0.5cm
\noindent{\bf Abstract.}\\
 We find that in the presence of an
electric field there is an attractive intergrain interaction in
superconductors which is small. When charge on the ball is permitted
to vary with the ball radius, very large balls can be formed.
The pairing energy makes the ball compact
and hence reduces the size of the ball compared with the
classical value. The ball radius depends on the gap of the superconductor
due to Josephson tunneling.
\vskip10.0cm
*Tel. +91-40-3010811; fax: +91-40-3010145\\
E-mail address: knssp@uohyd.ernet.in (K.N.Shrivastava)
\newpage
\noindent {\bf 1.Introduction.}\\
Some time ago, it was found that application of an
electric field produced a second-order interaction which led to
anisotropic resistivity with respect to a change in the sign of
the electric field [1]. Hence the change in sign of the applied
voltage which can be achieved by reversing the battery, resulted
into two different values of the resistivity for two different
polarities [2]. The geometry of the experimental configuration
for the application of the electric field plays an important
role. Therefore, another measurement was carried out by Frey et
al [3]. The second-order term in the potential due to the dipole
moment induced by the electric field leads to a resistivity
which is linear in electric field. At small values of the
electric field, $E\leq 0.3 E_{BD}$, this prediction [4] is in
accord with the experimental measurement. Here $E_{BD}$ is the
break-down field. The break-down voltage is $V_{BD}=E_{BD}d$ with
$d$ as the width of the film along the $c$ direction. At larger
values of the applied voltage, $0.16<E/E_{BD}<0.89$, the
resistivity is found [4] to depend on the square root of the
applied electric field, $R\propto E^{1/2}$. It is found [4,5] that
the density of states depend on the dimensionality so that
one-dimensional conduction normal to the surface of the film
produces the resistivity proportional to the square root of the
electric field. The same result is obtained by more elementary
considerations of the Thomas-Fermi screening length.

In the present paper, we show that there is an intergrain
interaction due to the electric dipole moment which is
attractive so that the grains tend to aggregate to make a ball.
The Josephson tunneling plays an important role to determine the
superconducting surface tension. We find that pairing
interaction helps in making a compact ball from superconducting
grains. The binding of Cooper pairs here, is the same as in the
B.C.S. theory except that large grains bind into a big ball.\\
\vskip0.5cm
{\bf 2. Theory}\\

We assume that the electron coordinates are $r_i$ and the charge
is $-e$ so that the dipole moment of one ion is $p=-e.r_i$.
Since different electrons are located at different coordinates
in an ion, we replace the dipole moment by the average value,
$<p>$. The potential energy of a grain due to dipole moment of
all of the ions within a grain is given by,
\beg
V_g^{(1)} = \sum_{i=1}^N V^{(1)} = -\sum_{i=1}^N E <p_i>
\eeq
where the sum is over all ions within a grain and $N$ is the
number of ions within the grain. In the case of two grains,
\beg
\sum_iV^{(1}) = - \sum^{N(1)}_{i=1} E<p_{i1}> -
\sum^{N(2)}_{j=1} E<p_{j2}>
\eeq
is the potential energy where $N(1)$ is the number of electrons
within the first grain and $N(2)$ is the number of electrons
within the second grain. The states of the grains are given by
$|0_2,0_1>$, $|n_2,n_1>$, etc. the second-order energy of the
system of two grains is,
\beg
V^{(2)} = \sum_i\sum_j \left[<n_1,n_2|Ep_{i1}|0_1,n_2><0_1,n_2|
Ep_{j2}|0_2,0_1>\over {E(0_1,n_2)-E(0_2,0_1)}\right]
\eeq
with one more term in which the subscripts 1 and 2 are
interchanged. This interaction varies as $E^2$ and attracts two
grains. Since $E$ is the electric field, it can be expanded into
photon operators. Then it means that one grain emits a photon
which is absorbed by other and vice versa. This interaction
travels with the speed of light and grains become attractive
except that it is small compared with the electromagnetic energy
$E^2/8\pi$.

The attraction between grains helps grains to aggregate together,
but is not sufficient for the ball formation. As proved by theoretical
calculations and numerous experiments, the dipolar attraction only
leads to form chains and columns along the field direction, never
leads to macroscopic balls (see reference 7). Therefore, the ball
formation reveals new and deep physics. The surface tension is
 $\sigma$ so that the surface
energy of a ball of radius $a$ is $4\pi a^2\sigma$. The ball is
near an electrode of charge $q$. The charge on the ball is $q$
and the dielectric constant of the medium is $\epsilon_o$. The
Coulomb energy is then $q^2/2a\epsilon_o$ and the
electromagnetic energy is $-3(4\pi
a^3/3)\epsilon_oE^2/8\pi=-(1/2)\epsilon_oE^2a^3$. The energy of
the ball is then given by,
\beg
U = 4\pi a^2\sigma + q^2/(2\epsilon_oa)-(1/2)\epsilon_oE^2a^3.
\eeq
{\it Case I.} We discuss two cases of this free energy separately.
For constant q or when charge is independent of the size, the above is
 minimized with respect to the radius of the ball by
setting $dU/da=0$ so that,
\beg
\sigma = q^2/(16\pi\epsilon_oa^3) + 3E^2a\epsilon_o/(16\pi)\,\,\,.
\eeq
The charge, $q$, of the ball is proportional to
$E\epsilon_oa^2$. Therefore we assume that $q=\gamma
E\epsilon_oa^2$ which substituted in the above gives a relation
between applied electric field $E$ and the radius of the ball as,
\beg
E^2a = 16\pi\sigma/[\epsilon_o(3+\gamma^2)]
\eeq
so that the radius of the ball depends on the inverse square of
the electric field, $a\propto1/E^2$. This relation seems to be
marginally satisfactory for NdBa$_2$Cu$_3$O$_{7-\delta}$ but in
the case of Bi$_2$Sr$_2$CaCu$_2$O$_{8+\delta}$, the relation is
obeyed only for large values of $E$. For $E<0.94$ kV/mm there is
a deviation between the measured values and those calculated from
$E^2a=$ constant. The measured [7] values of the radius of the
ball are smaller than those calculated. None of the three terms
in (4) are explicitly dependent on quantum effects. The surface
energy is the product of the surface area $4\pi a^2$ and the
surface tension but does not have explicit dependence on quantum
nature of superconductivity. The Coulomb energy is just the
square of the charge devided by the distance and the
electrostatic energy also does not involve any quantum effects.
If the electrostatic energy was dominant, there will be crystal
growth according to the crystallographic symmetry. Since the
growth is spherical, according to the crystal symmetries, there
will be a texture which minimizes the energy.\\
{\it Case II.} We consider that in the free energy
(4) the charge depends on the radius of the ball, $a$.
 Therefore,
we treat the charge as dependent on the radius. The charge
$q=\gamma E\epsilon_oa^2$ is eliminated from (4) so that the
free energy becomes,
\beg
U=4\pi a^2\sigma + {\gamma^2E^2\epsilon_oa^3\over2} -
\left({1\over2}\right) \epsilon_oE^2a^3\,\,\,.
\eeq
Minimizing this free energy with respect to the radius of the
ball we set $dU/da=0$ which gives,
\beg
E^2a = {16\pi\sigma\over3\epsilon_o(1-\gamma^2)}.
\eeq
This gives maximum, if $\gamma^2\ll 1$. In fact, the minimum is at
$a=\infty$ when $\gamma^2\ll 1$. This will lead to the formation
of large balls. If $\gamma^2\gg 1$, the minimum is at $a=0$ so that balls
can not be formed.\\

{\it Case III.} Pairing with constant charge.
We introduce the pairing energy which is important for
superconductivity and explains the experimentally measured
values of the radius of the superconducting ball. The
electron-phonon interaction is given by $\sum_{k,k^\prime}D
c^\dagger_{k,\sigma}c_{k^\prime,\sigma}a_{k-k^\prime}+h.c.$
where $h.c.$ stands for the hermitian conjugate of the previous
term. The
$c^\dagger_{k,\sigma}(c_{k,\sigma})$ are the creation
(annihilation) operators for electrons of wave vector {\bf k}
and spin $\sigma$ and $a^\dagger_q(a_q)$ for phonons of wave
vector {\bf q} and frequency $\omega$. Here $D$ is the
interaction constant. This interaction leads to the attractive potential,
\beg
V_{eff} = {2D^2\hbar\omega_q
c^\dagger_{k\uparrow}c^\dagger_{k\downarrow} c_{k\downarrow}c_{k\uparrow}\over
(\epsilon_k-\epsilon_{k^\prime})^2-(\hbar\omega_q)^2}
\eeq
where $\hbar\omega_q$ is the phonon single-particle energy and
$\epsilon_k$ are the electron single-particle energies. The
attractive interaction is achieved when
$\epsilon_k-\epsilon_{k^\prime}\ll\hbar\omega_q$. This
condition introduces the negative sign with respect to the
kinetic energy terms. The pairing operators are averaged as
$<c^\dagger_{k^\uparrow}c^\dagger_{k\downarrow}>=<c_{k\downarrow}c_{k\uparrow}>
= {\Delta\over V}$ where $V$ is the attractive potential
$V=-2D^2/(\hbar\omega_q)$. The kinetic energy terms of the
hamiltonian are slightly renormalized in going from the normal
to the superconducting state. However ignoring this small
renormalization effect, we can write the average value of the
B.C.S. hamiltonian as $-2\Delta^2/V=\Delta^2\hbar\omega/D^2$ where
$\Delta$ is the gap energy. Therefore eq.(4) is subject to a
quantum correction due to pairing energy. The volume of the ball
is $4\pi a^3/3$ and that of a Cooper pair is $4\pi\xi^3/3$ so
that the number of Cooper pairs is $(a/\xi)^3$. Since $\Delta$
is the gap in the single-particle dispersion relation the
pairing energy inside the superconducting ball is
$(\Delta^2\hbar\omega/D^2)(a/\xi)^3$. Therefore the energy of the
ball becomes,
\beg
U = 4\pi\sigma a^2 + q^2/(2\epsilon_oa) -
a^3[\Delta^2\hbar\omega/(D^2\xi^3)+(1/2) E^2\epsilon_o]\,\,.
\eeq
Minimizing $U$ with respect to the radius of the ball, we set
$dU/da=0$ which we solve for the surface tension to find,
\beg
\sigma = {q^2\over16\pi\epsilon_oa^3} + {3a\Delta^2\hbar\omega\over
8\pi D^2\xi^3} + {3aE^2\epsilon_o\over16\pi}\,\,\,,
\eeq
for a constant charge on the ball. Substituting the charge of
the ball $q=\gamma E\epsilon_oa^2$ in eq.(11) and solving for
the radius we obtain,
\beg
E^2a = 16\pi\sigma/[\epsilon_o\{3+\gamma^2+a_1\}]
\eeq
where
\beg
a_1 = 6\Delta^2\hbar\omega/[\epsilon_oD^2E^2\xi^3] =
a_2(\Delta^2/\xi^3) \,\,.
\eeq
When $E$ is reduced, $a_1$ increases and hence the denominator
in (12) increases and $a$ reduces. We assume that $a_1$ is a
small number, $a_1/(3+\gamma^2)\ll1$. The eq.(12) then can be
written by using the binomial theorem expansion and retaining
only the first two terms as,
\beg
E^2a =
[16\pi\sigma/\{\epsilon_o(3+\gamma^2)\}]\{1-a_1/(3+\gamma^2)\} \,\,.
\eeq
The radius of the superconducting ball given by this expression
is smaller than that given by (6) due to pairing of electrons. The
classical ball is thus compressed by the pairing energy. Since
the mass of the ball is independent of the pairing of electrons
and the volume is reduced we find that the density of the ball
increases due to pair formation. In the case of strong pairing,
$a_1\gg 3+\gamma^2$, the eq.(12) gives,
\beg
a = 8\pi\sigma D^2\xi^3/(3\Delta^2\hbar\omega)
\eeq
and the radius of the superconducting ball becomes independent
of the applied electric field. This means that when charge on
the ball is small and the tunneling current is small, then the
ball can move in the electric field with a constant radius. It
is also clear that when the surface tension vanishes,
$\sigma=0$, the ball collapses with zero radius, $a=0$.

The surface tension on the superconducting ball is caused by the
Josephson tunneling along the surface of the ball which means
that the $c$ axis is tangential to the radius so that
\beg
\sigma^\prime = Jc_o
\eeq
where $c_o$ is the unit cell dimension along the $c$ axis and
the Josephson coupling energy is 
\beg
{\cal H}^\prime = - J\cos\theta
\eeq
where
\beg
\theta = \theta_1 - \theta_2 - (2e/hc) \int \mbox{{\bf A}.d{\em l}}
\eeq
is the phase factor. All the grains are aligned in such a way
that the $c$ axis is always on the surface. Keeping the $c$ axis
on the surface can be achieved by rotations along the $a$ or $b$
axis which are equivalent. Therefore the superconducting ball
develops a texture.

It was found by us [6] that the normal effects can be changed to
superconducting properties by introducing the factor of
$l_s/\xi$ where $l_s$ is the mean free path of normal electrons
and $\xi$ is the coherence length. The superconducting ball gets
charged by the electrodes which is a normal effect. Therefore,
due to this normal charge the superconducting Josephson current
is reduced by the factor $l_s/\xi$. We suppose that $n$ is the
concentration of normal electrons so that the mean free path is
$l_s=(\pi/3)^{1/6}[a_o/(4n^{1/3})]^{1/2}$ with
$a_o=\hbar^2/me^2$ as the Bohr radius. Hence, we can write the
surface tension on the surface of the charged ball as,
\beg
\sigma = J c_ol_s/\xi\,\,\,,
\eeq
which replaces (16). It is sufficient for the present purpose to
write $J\approx J_o \approx \pi\Delta/2SR_N$ where $S$ is the
surface area and $R_N$ is the normal resistivity and
$\Delta=\Delta_o[1-T/T_c]^{1/2}$ so that we can estimate the
temperature dependence of the ball radius from,
\beg
a = [\pi c_ol_sD^2\xi^2/(3\Delta\hbar\omega R_N)]^{1/3}
\eeq
where we have used (15), (19) and $S=4\pi a^2$. Using the fact
that $\xi$ diverges as $\xi=\xi_o/(1-T/T_c)^\nu$ where
$\nu\approx0.7$ is the exponent for the divergence of the
coherence length, the above expression can be written as,
\beg
a=[\pi c_ol_sD^2\xi_o^2/(3\hbar\omega
R_N\Delta_o)]^{1/3}/(1-T/T_c)^{1/6+2\nu/{3}} \,\,.
\eeq
At $T=T_c$, the coherence length diverges and hence, the ball
radius for strong pairing shows strong divergence.\\
{\it Case IV.} As noted in
(7), for the charge depending on radius we rewrite the eq.(10) as
\beg
U = 4\pi\sigma a^2 + {\gamma^2E^2\epsilon_oa^3\over2} -
a^3[\Delta^2\hbar\omega/(D^2\xi^3)+ ({1\over2}) E^2\epsilon_o]
\eeq
which gives results similar to those already discussed.

A scanning electron micrograph of the superconducting ball
formed by the application of an electric field on a powder of
superconducting material contains small grains dispersed in
liquid nitrogen is shown by Tao et al [7]. It is quite clear
that the ball is not perfectly spherical. If the pairing of
electrons is an $s$-wave type, we would expect the formation of
a perfectly spherical ball. Therefore, we think that the gap has
$d$-wave symmetry. In which case the gap in (10) and in
subsequent relations should be replaced by
$\Delta=\Delta_o\cos2\varphi$. It is also possible that the gap
is of complex nature in which the symmetry changes from
$d(x^2-y^2)$ to $d(xy)$ or from $s$ to $d(x^2-y^2)$ when
temperature or magnetic field is varied. The high temperature
phase may have higher symmetry than the low temperature phase as
found earlier [8]. A detailed study of the effect of the electric
field on superconductors is given in a recent book[9].\\
{\bf 3.Conclusions.}\\

In conclusion, we find that superconducting powder forms a ball
when subjected to an electric field. The pairing interaction
plays an important role while Josephson interaction provides the
surface tension. The temperature dependence of the ball radius
arises from the divergence in the coherence length.
\vskip0.1cm
{\bf References.}
\noindent{}[1] K.N. Shrivastava, J. Phys.: Condens. Matter {\bf5} (1993) L597.\\
{}[2] J. Mannhart, D.G. Schlom, J.G. Bednorz and K.A. M\"uller,
	Phys. Rev. Lett. {\bf67} (1991) 2099.\\
{}[3] T. Frey, J. Mannhart, J.G. Bednorz and E.J. Williams,
	Phys. Rev. B{\bf51} (1995) 3257.\\
{}[4] L.S. Lingam and K.N. Shrivastava, Mod. Phys. Lett.
	B{\bf10} (1996) 1123.\\
{}[5] L.S. Lingam and K. N. Shrivastava, Physica B {\bf223}(1996)577.\\ 
{}[6] K.N.Shrivastava, J. Phys.(Paris)Colloq {\bf49}(1988)C8-2239.\\
{}[7] R. Tao, X. Zhang, X. Tang and P.W. Anderson, Phys. Rev.
	Lett. {\bf83} (1999) 5575.\\
{}[8] N.M. Krishna and K.N. Shrivastava, Physica B{\bf230}
(1997) 939.\\
{}[9] K. N. Shrivastava, Superconductivity:Elementary Topics, World
      Scientific Pub., New Jersey, London, Hong Kong, Singapore, (2000).\\

\end{document}